\documentclass[10pt,a4paper]{article}

\usepackage[breaklinks]{hyperref}
\usepackage[T1]{fontenc} 
\usepackage[latin1]{inputenc}

\usepackage{aeguill}
\usepackage{tikz}
\usepackage{amsfonts}
\usepackage{amsmath}
\usepackage{amssymb}
\usepackage{amssymb,stmaryrd}
\usepackage{gnuplot-lua-tikz}
\usepackage{subfigure}
\usepackage{multirow,multicol}
\usepackage{garamond}
\usepackage[sort&compress]{natbib}
\usepackage[]{url,hyperref}

\usetikzlibrary{arrows,automata,shapes}

\newtheorem{definition}{Definition}

\begin{document}

\garamond

\title{Multi-level agent-based modeling\\~\\A literature survey\\~\\}

\author{Gildas Morvan\\~\\
{\small \url{http://www.lgi2a.univ-artois.fr/~morvan/}}\\{\small \url{gildas.morvan@univ-artois.fr}}\\~\\Univ Lille Nord de France, F-59000 Lille, France\\UArtois, LGI2A, F-62400, Béthune, France\\~\\
\includegraphics[width=3cm]{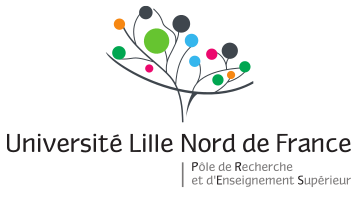}~~~\includegraphics[width=3cm]{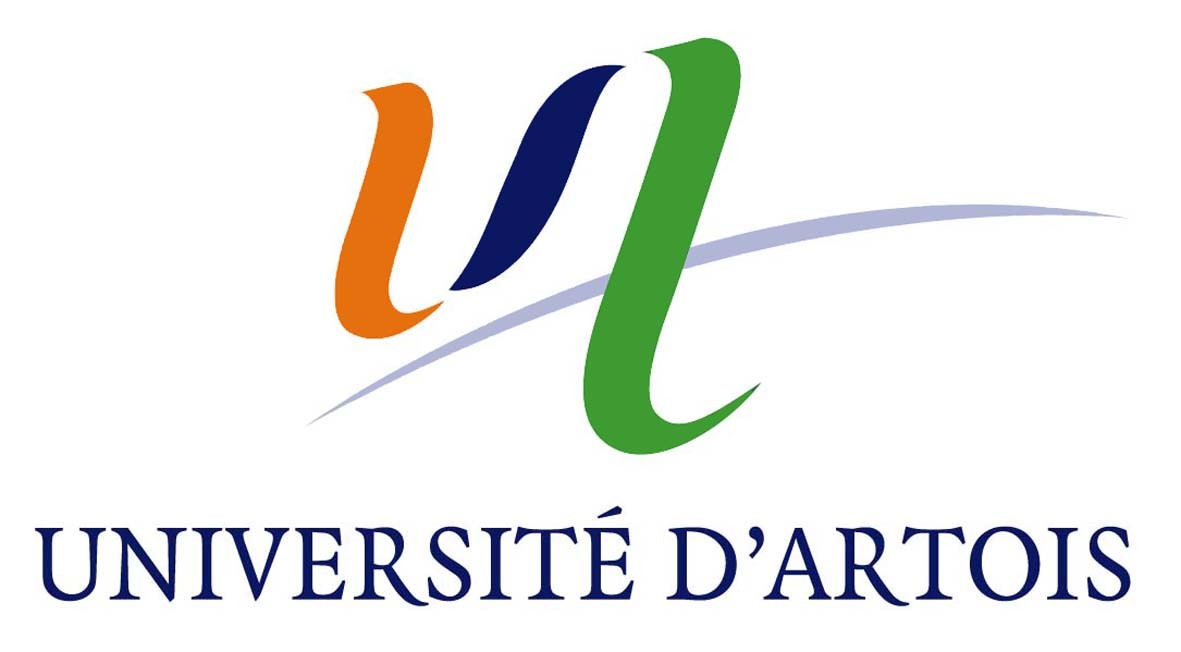}~~~\includegraphics[width=2.5cm]{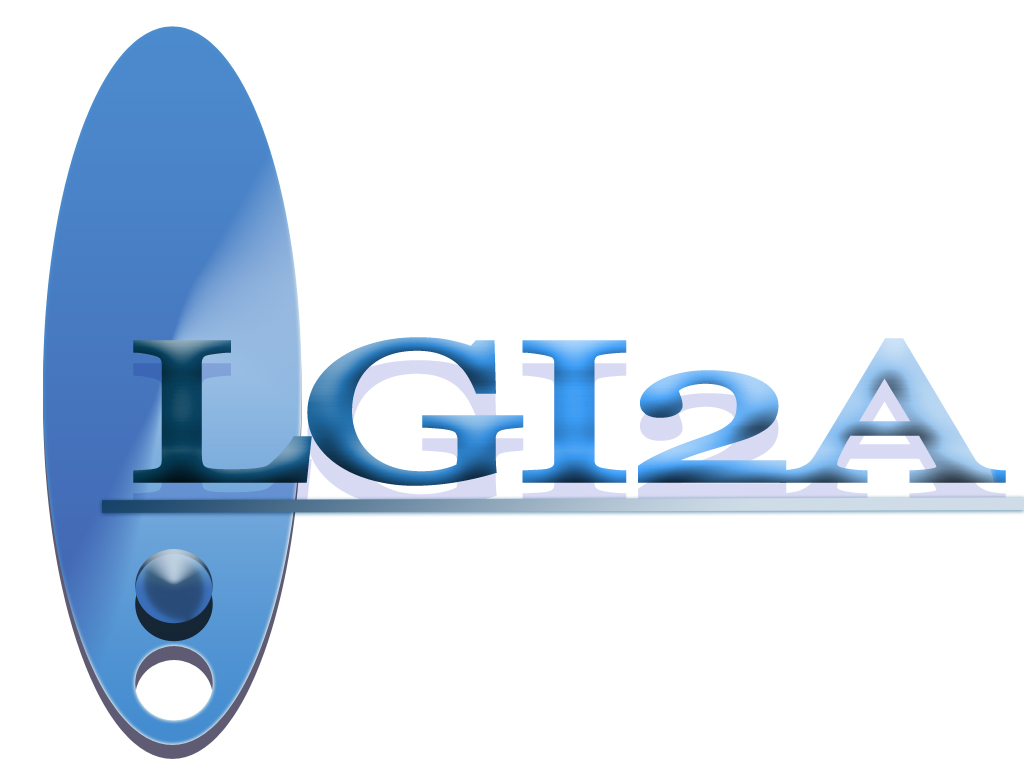}\\~\\
}

\maketitle

\setcounter{footnote}{0}

\pagebreak

\section*{Absract}
During last decade, multi-level agent-based modeling has received significant and dramatically increasing interest. In this article we conduct a comprehensive and structured review of literature on this emerging research domain that aims at extending the classical agent-based modeling paradigm to overcome some of its limitations. We present the main theoretical contributions and applications  with an emphasis on social, flow, biological and biomedical models.

\section*{Main acronyms used in this paper}

\paragraph{ABM} According to the context, Agent-Based Model or Agent-Based Modeling
\paragraph{DEVS} Discrete Event System Specification
\paragraph{EBM} Equation-Based Model
\paragraph{ML-ABM} According to the context, Multi-Level Agent-Based Model or Multi-Level Agent-Based Modeling

\setcounter{tocdepth}{2}
 \tableofcontents

\pagebreak

\section{Introduction}
\label{Introduction}

\subsection{Agent-based modeling}

Agent-based modeling (ABM) is a computational modeling paradigm that allow to simulate the interactions of autonomous agents in an environment. It has been widely used to study complex systems in various domains~\citep{Ferber:1999,Epstein:2006,Gilbert:2007,Railsback:2011,Resnick:1994,Treuil:2008}.

However, it suffers from important known limitations that reduce its scope~\citep{Drogoul:2003a,Scerri:2010}.
First, ABM is purely bottom-up: a microscopic knowledge, \textit{i.e.}, related to system components, is used to construct models while a macroscopic  knowledge, \textit{i.e.}, related to global system properties, is used to validate models.Therefore,  it is not straightforward to explicitly introduce bidirectional relations between these two points of view in the general case. It becomes even harder when different spatial or temporal scales or domains of interest are involved in a same simulation.
Moreover, agent-based models do not scale easily and generally require large computational resources since many agents are simulated.
Finally, most agent-based simulation platforms lack tools to reify complex singular emergent properties: human observation often remains the most efficient way to capture multi-level pattern formation or crowd behavior.



\subsection{Multi-level agent-based modeling}
\label{mlabm}

\subsubsection{Definitions}

The works surveyed in this paper aim at extending the classical ABM paradigm to overcome these limitations. While they can be very different (in terms of goals, technical approaches or application domains), these woks share a common idea: introducing more levels of description in agent-based models. Therefore, we group them under a common term: Multi-Level Agent-based Modeling (ML-ABM). In the following we propose definitions of the concepts of level and multi-level agent-based model that seem broad enough to encompass the different surveyed approaches~(see section~\ref{Terminology}).

\begin{definition}
	A level is a point of view on a system, integrated in a model as a specific abstraction.
	\label{def:level}
\end{definition}

The name refers to the familiar expressions levels of organization, observation, analysis, etc.

\begin{definition}
	A multi-level agent-based model \textbf{integrates} \textbf{heterogenous} (agent-based) models, representing \textbf{complementary points of view}, so called levels, of the same system.
	\label{def:mlabm}
\end{definition}

In this definition, three concepts are highlighted and should themselves be defined formally.

\begin{definition}
\textbf{Integration} means that the ABMs within a ML-ABM can interact and share entities such as environments and agents.
\end{definition}
%

\begin{definition}
\textbf{Heterogeneity} means that the ABMs integrated in a ML-ABM can be based on different modeling paradigms (differential equations, cellular automata, etc.), use different time representation (discrete events, step-wise) and represent processes at different spatio-temporal scales.
\end{definition}
\begin{definition}
 \textbf{Points of view} are \textbf{complementary} for a given problem since they can not be taken in isolation to address it. 
\end{definition}

This idea is very important in the literature on complex systems~\citep{Morin:1992}. Indeed, as~\citet{Muller:2011} note, "the global behavior of a complex system cannot be understood without letting a set of points of view interact".

\subsubsection{Types of problems solved using multi-level agent-based approaches }

ML-ABM  is mainly used to solve three types of \textit{modeling} problems:
\begin{itemize}
\item the modeling of \textbf{cross-level interactions}, \textit{e.g.}, an explicit top-down feedback control,
\item the \textbf{coupling of heterogeneous models},
\item the \textbf{(dynamic) adaptation of the level of detail} of simulations, \textit{e.g.}, to save computational resources or use the best available model in a given context.
\end{itemize}
In the first case, the different points of view always co-exist, as they integrate interdependent models,  while in the last ones, levels are (de)activated at run-time according to the context, as they represent independent models designed for specific situations.

For instance, in flow hybrid models areas with simple topologies are handled with an equation-based model (EBM) while others are handled with an ABM.

\subsection{Terminology issues}
\label{Terminology}

Different terms have been used to describe what we call here a level: \textit{e.g.}, perspective, interaction, layer  or view~\citep{Desmeulles:2009,White:2007,Torii:2005,Parunak:2009}. Some are domain-specific; thus, in the flow modeling domain, when two levels with static relations are considered, models are often  described as \textit{hybrid} as shown in the section~\ref{Flow}~\citep{Burghout:2005,El-hmam:2006,El-hmam:2006a,El-Hmam:2006b,El-hmam:2008,Marino:2011,Mathieu:2007a,Nguyen:2012,Rejniak:2011,Wakeland:2007}.

The term multi-scale ---~or multi-resolution~\citep{Jeschke:2008,Zhang:2009a,Zhang:2011}~--- is often used but has a more restrictive meaning as it focuses on the spatial and temporal extents of levels and not on their interactions and organization. \citet[p. 622--623]{Gil-Quijano:2012} pointed that the term multi-scale can be misleading and advocated for using multi-level instead. Let take as an example the \textit{Simpop3} model, described by their authors as multi-scale~\citep{Pumain:2009a}. Two levels are considered: the \textit{city level}, representing the internal dynamics of a city, and the  \textit{system of cities level}, representing the interactions between cities. However, the idea of scale does not fit to describe the relation between them: one can easily figure a city bigger (in terms of population, spatial extent, economic exchanges, etc.) than a system of cities. In contrast, the idea of levels of organization in interaction seems more appropriate. Furthermore, to extend the definitions \ref{def:level} and \ref{def:mlabm} such models could be more precisely denoted as nested or hierarchical multi-level agent-based models.

\subsection{Bibliography}

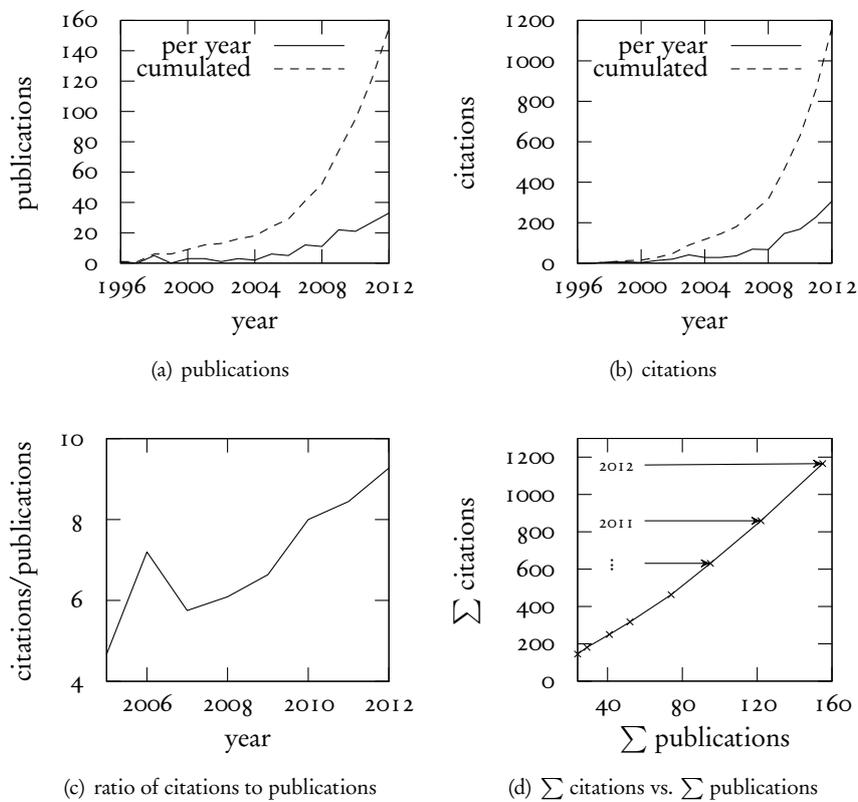
\begin{figure}[h]
	\begin{center}
		\subfigure[publications]{\begin{tikzpicture}[gnuplot]
\gpmonochromelines
\path (0.000,0.000) rectangle (5.588,4.572);
\gpcolor{color=gp lt color border}
\gpsetlinetype{gp lt border}
\gpsetlinewidth{1.00}
\draw[gp path] (1.504,0.985)--(1.684,0.985);
\draw[gp path] (5.035,0.985)--(4.855,0.985);
\node[gp node right] at (1.320,0.985) { 0};
\draw[gp path] (1.504,1.387)--(1.684,1.387);
\draw[gp path] (5.035,1.387)--(4.855,1.387);
\node[gp node right] at (1.320,1.387) { 20};
\draw[gp path] (1.504,1.790)--(1.684,1.790);
\draw[gp path] (5.035,1.790)--(4.855,1.790);
\node[gp node right] at (1.320,1.790) { 40};
\draw[gp path] (1.504,2.192)--(1.684,2.192);
\draw[gp path] (5.035,2.192)--(4.855,2.192);
\node[gp node right] at (1.320,2.192) { 60};
\draw[gp path] (1.504,2.594)--(1.684,2.594);
\draw[gp path] (5.035,2.594)--(4.855,2.594);
\node[gp node right] at (1.320,2.594) { 80};
\draw[gp path] (1.504,2.996)--(1.684,2.996);
\draw[gp path] (5.035,2.996)--(4.855,2.996);
\node[gp node right] at (1.320,2.996) { 100};
\draw[gp path] (1.504,3.399)--(1.684,3.399);
\draw[gp path] (5.035,3.399)--(4.855,3.399);
\node[gp node right] at (1.320,3.399) { 120};
\draw[gp path] (1.504,3.801)--(1.684,3.801);
\draw[gp path] (5.035,3.801)--(4.855,3.801);
\node[gp node right] at (1.320,3.801) { 140};
\draw[gp path] (1.504,4.203)--(1.684,4.203);
\draw[gp path] (5.035,4.203)--(4.855,4.203);
\node[gp node right] at (1.320,4.203) { 160};
\draw[gp path] (1.504,0.985)--(1.504,1.165);
\draw[gp path] (1.504,4.203)--(1.504,4.023);
\node[gp node center] at (1.504,0.677) { 1996};
\draw[gp path] (2.387,0.985)--(2.387,1.165);
\draw[gp path] (2.387,4.203)--(2.387,4.023);
\node[gp node center] at (2.387,0.677) { 2000};
\draw[gp path] (3.270,0.985)--(3.270,1.165);
\draw[gp path] (3.270,4.203)--(3.270,4.023);
\node[gp node center] at (3.270,0.677) { 2004};
\draw[gp path] (4.152,0.985)--(4.152,1.165);
\draw[gp path] (4.152,4.203)--(4.152,4.023);
\node[gp node center] at (4.152,0.677) { 2008};
\draw[gp path] (5.035,0.985)--(5.035,1.165);
\draw[gp path] (5.035,4.203)--(5.035,4.023);
\node[gp node center] at (5.035,0.677) { 2012};
\draw[gp path] (1.504,4.203)--(1.504,0.985)--(5.035,0.985)--(5.035,4.203)--cycle;
\node[gp node center,rotate=-270] at (0.246,2.594) {publications};
\node[gp node center] at (3.269,0.215) {year};
\node[gp node right] at (3.344,3.869) {per year};
\gpcolor{color=gp lt color 0}
\gpsetlinetype{gp lt plot 0}
\draw[gp path] (3.528,3.869)--(4.444,3.869);
\draw[gp path] (1.504,1.005)--(1.725,0.985)--(1.945,1.086)--(2.166,0.985)--(2.387,1.045)%
  --(2.607,1.045)--(2.828,1.005)--(3.049,1.045)--(3.270,1.025)--(3.490,1.106)--(3.711,1.086)%
  --(3.932,1.226)--(4.152,1.206)--(4.373,1.427)--(4.594,1.407)--(4.814,1.528)--(5.035,1.649);
\gpcolor{color=gp lt color border}
\node[gp node right] at (3.344,3.561) {cumulated};
\gpcolor{color=gp lt color 1}
\gpsetlinetype{gp lt plot 1}
\draw[gp path] (3.528,3.561)--(4.444,3.561);
\draw[gp path] (1.504,1.005)--(1.725,1.005)--(1.945,1.106)--(2.166,1.106)--(2.387,1.166)%
  --(2.607,1.226)--(2.828,1.246)--(3.049,1.307)--(3.270,1.347)--(3.490,1.468)--(3.711,1.568)%
  --(3.932,1.810)--(4.152,2.031)--(4.373,2.473)--(4.594,2.896)--(4.814,3.439)--(5.035,4.102);
\gpcolor{color=gp lt color border}
\gpsetlinetype{gp lt border}
\draw[gp path] (1.504,4.203)--(1.504,0.985)--(5.035,0.985)--(5.035,4.203)--cycle;
\gpdefrectangularnode{gp plot 1}{\pgfpoint{1.504cm}{0.985cm}}{\pgfpoint{5.035cm}{4.203cm}}
\end{tikzpicture}
}
		\subfigure[citations]{\begin{tikzpicture}[gnuplot]
\gpmonochromelines
\path (0.000,0.000) rectangle (5.588,4.572);
\gpcolor{color=gp lt color border}
\gpsetlinetype{gp lt border}
\gpsetlinewidth{1.00}
\draw[gp path] (1.688,0.985)--(1.868,0.985);
\draw[gp path] (5.035,0.985)--(4.855,0.985);
\node[gp node right] at (1.504,0.985) { 0};
\draw[gp path] (1.688,1.521)--(1.868,1.521);
\draw[gp path] (5.035,1.521)--(4.855,1.521);
\node[gp node right] at (1.504,1.521) { 200};
\draw[gp path] (1.688,2.058)--(1.868,2.058);
\draw[gp path] (5.035,2.058)--(4.855,2.058);
\node[gp node right] at (1.504,2.058) { 400};
\draw[gp path] (1.688,2.594)--(1.868,2.594);
\draw[gp path] (5.035,2.594)--(4.855,2.594);
\node[gp node right] at (1.504,2.594) { 600};
\draw[gp path] (1.688,3.130)--(1.868,3.130);
\draw[gp path] (5.035,3.130)--(4.855,3.130);
\node[gp node right] at (1.504,3.130) { 800};
\draw[gp path] (1.688,3.667)--(1.868,3.667);
\draw[gp path] (5.035,3.667)--(4.855,3.667);
\node[gp node right] at (1.504,3.667) { 1000};
\draw[gp path] (1.688,4.203)--(1.868,4.203);
\draw[gp path] (5.035,4.203)--(4.855,4.203);
\node[gp node right] at (1.504,4.203) { 1200};
\draw[gp path] (1.688,0.985)--(1.688,1.165);
\draw[gp path] (1.688,4.203)--(1.688,4.023);
\node[gp node center] at (1.688,0.677) { 1996};
\draw[gp path] (2.525,0.985)--(2.525,1.165);
\draw[gp path] (2.525,4.203)--(2.525,4.023);
\node[gp node center] at (2.525,0.677) { 2000};
\draw[gp path] (3.362,0.985)--(3.362,1.165);
\draw[gp path] (3.362,4.203)--(3.362,4.023);
\node[gp node center] at (3.362,0.677) { 2004};
\draw[gp path] (4.198,0.985)--(4.198,1.165);
\draw[gp path] (4.198,4.203)--(4.198,4.023);
\node[gp node center] at (4.198,0.677) { 2008};
\draw[gp path] (5.035,0.985)--(5.035,1.165);
\draw[gp path] (5.035,4.203)--(5.035,4.023);
\node[gp node center] at (5.035,0.677) { 2012};
\draw[gp path] (1.688,4.203)--(1.688,0.985)--(5.035,0.985)--(5.035,4.203)--cycle;
\node[gp node center,rotate=-270] at (0.246,2.594) {citations};
\node[gp node center] at (3.361,0.215) {year};
\node[gp node right] at (3.528,3.869) {per year};
\gpcolor{color=gp lt color 0}
\gpsetlinetype{gp lt plot 0}
\draw[gp path] (3.712,3.869)--(4.628,3.869);
\draw[gp path] (1.688,0.985)--(1.897,0.985)--(2.106,1.001)--(2.316,1.001)--(2.525,0.993)%
  --(2.734,1.020)--(2.943,1.039)--(3.152,1.095)--(3.362,1.060)--(3.571,1.060)--(3.780,1.082)%
  --(3.989,1.170)--(4.198,1.165)--(4.407,1.377)--(4.617,1.436)--(4.826,1.596)--(5.035,1.806);
\gpcolor{color=gp lt color border}
\node[gp node right] at (3.528,3.561) {cumulated};
\gpcolor{color=gp lt color 1}
\gpsetlinetype{gp lt plot 1}
\draw[gp path] (3.712,3.561)--(4.628,3.561);
\draw[gp path] (1.688,0.985)--(1.897,0.985)--(2.106,1.001)--(2.316,1.017)--(2.525,1.025)%
  --(2.734,1.060)--(2.943,1.114)--(3.152,1.224)--(3.362,1.299)--(3.571,1.374)--(3.780,1.470)%
  --(3.989,1.655)--(4.198,1.835)--(4.407,2.227)--(4.617,2.677)--(4.826,3.289)--(5.035,4.109);
\gpcolor{color=gp lt color border}
\gpsetlinetype{gp lt border}
\draw[gp path] (1.688,4.203)--(1.688,0.985)--(5.035,0.985)--(5.035,4.203)--cycle;
\gpdefrectangularnode{gp plot 1}{\pgfpoint{1.688cm}{0.985cm}}{\pgfpoint{5.035cm}{4.203cm}}
\end{tikzpicture}
}
		\subfigure[ratio of citations to publications]{\begin{tikzpicture}[gnuplot]
\gpmonochromelines
\path (0.000,0.000) rectangle (5.588,4.572);
\gpcolor{color=gp lt color border}
\gpsetlinetype{gp lt border}
\gpsetlinewidth{1.00}
\draw[gp path] (1.320,0.985)--(1.500,0.985);
\draw[gp path] (5.035,0.985)--(4.855,0.985);
\node[gp node right] at (1.136,0.985) { 4};
\draw[gp path] (1.320,2.058)--(1.500,2.058);
\draw[gp path] (5.035,2.058)--(4.855,2.058);
\node[gp node right] at (1.136,2.058) { 6};
\draw[gp path] (1.320,3.130)--(1.500,3.130);
\draw[gp path] (5.035,3.130)--(4.855,3.130);
\node[gp node right] at (1.136,3.130) { 8};
\draw[gp path] (1.320,4.203)--(1.500,4.203);
\draw[gp path] (5.035,4.203)--(4.855,4.203);
\node[gp node right] at (1.136,4.203) { 10};
\draw[gp path] (1.851,0.985)--(1.851,1.165);
\draw[gp path] (1.851,4.203)--(1.851,4.023);
\node[gp node center] at (1.851,0.677) { 2006};
\draw[gp path] (2.912,0.985)--(2.912,1.165);
\draw[gp path] (2.912,4.203)--(2.912,4.023);
\node[gp node center] at (2.912,0.677) { 2008};
\draw[gp path] (3.974,0.985)--(3.974,1.165);
\draw[gp path] (3.974,4.203)--(3.974,4.023);
\node[gp node center] at (3.974,0.677) { 2010};
\draw[gp path] (5.035,0.985)--(5.035,1.165);
\draw[gp path] (5.035,4.203)--(5.035,4.023);
\node[gp node center] at (5.035,0.677) { 2012};
\draw[gp path] (1.320,4.203)--(1.320,0.985)--(5.035,0.985)--(5.035,4.203)--cycle;
\node[gp node center,rotate=-270] at (0.246,2.594) {citations/publications};
\node[gp node center] at (3.177,0.215) {year};
\gpcolor{color=gp lt color 0}
\gpsetlinetype{gp lt plot 0}
\draw[gp path] (1.320,1.343)--(1.851,2.701)--(2.381,1.924)--(2.912,2.106)--(3.443,2.399)%
  --(3.974,3.130)--(4.504,3.369)--(5.035,3.813);
\gpcolor{color=gp lt color border}
\gpsetlinetype{gp lt border}
\draw[gp path] (1.320,4.203)--(1.320,0.985)--(5.035,0.985)--(5.035,4.203)--cycle;
\gpdefrectangularnode{gp plot 1}{\pgfpoint{1.320cm}{0.985cm}}{\pgfpoint{5.035cm}{4.203cm}}
\end{tikzpicture}
}
		\subfigure[$\sum$ citations vs. $\sum$ publications]{\begin{tikzpicture}[gnuplot]
\gpmonochromelines
\path (0.000,0.000) rectangle (5.588,4.572);
\gpcolor{color=gp lt color border}
\gpsetlinetype{gp lt border}
\gpsetlinewidth{1.00}
\draw[gp path] (1.688,0.985)--(1.868,0.985);
\draw[gp path] (5.035,0.985)--(4.855,0.985);
\node[gp node right] at (1.504,0.985) { 0};
\draw[gp path] (1.688,1.480)--(1.868,1.480);
\draw[gp path] (5.035,1.480)--(4.855,1.480);
\node[gp node right] at (1.504,1.480) { 200};
\draw[gp path] (1.688,1.975)--(1.868,1.975);
\draw[gp path] (5.035,1.975)--(4.855,1.975);
\node[gp node right] at (1.504,1.975) { 400};
\draw[gp path] (1.688,2.470)--(1.868,2.470);
\draw[gp path] (5.035,2.470)--(4.855,2.470);
\node[gp node right] at (1.504,2.470) { 600};
\draw[gp path] (1.688,2.965)--(1.868,2.965);
\draw[gp path] (5.035,2.965)--(4.855,2.965);
\node[gp node right] at (1.504,2.965) { 800};
\draw[gp path] (1.688,3.460)--(1.868,3.460);
\draw[gp path] (5.035,3.460)--(4.855,3.460);
\node[gp node right] at (1.504,3.460) { 1000};
\draw[gp path] (1.688,3.955)--(1.868,3.955);
\draw[gp path] (5.035,3.955)--(4.855,3.955);
\node[gp node right] at (1.504,3.955) { 1200};
\draw[gp path] (2.082,0.985)--(2.082,1.165);
\draw[gp path] (2.082,4.203)--(2.082,4.023);
\node[gp node center] at (2.082,0.677) { 40};
\draw[gp path] (3.066,0.985)--(3.066,1.165);
\draw[gp path] (3.066,4.203)--(3.066,4.023);
\node[gp node center] at (3.066,0.677) { 80};
\draw[gp path] (4.051,0.985)--(4.051,1.165);
\draw[gp path] (4.051,4.203)--(4.051,4.023);
\node[gp node center] at (4.051,0.677) { 120};
\draw[gp path] (5.035,0.985)--(5.035,1.165);
\draw[gp path] (5.035,4.203)--(5.035,4.023);
\node[gp node center] at (5.035,0.677) { 160};
\draw[gp path] (1.688,4.203)--(1.688,0.985)--(5.035,0.985)--(5.035,4.203)--cycle;
\node[gp node center,rotate=-270] at (0.246,2.594) {$\sum$ citations};
\node[gp node center] at (3.361,0.215) {$\sum$ publications};
\node[gp node left] at (1.836,3.869) {\scriptsize{2012}};
\node[gp node left] at (1.836,3.111) {\scriptsize{2011}};
\node[gp node center,rotate=-270] at (2.082,2.547) {\scriptsize{...}};
\draw[gp path,->](2.574,3.851)--(4.912,3.869);
\draw[gp path,->](2.574,3.111)--(4.100,3.111);
\draw[gp path,->](2.574,2.547)--(3.435,2.547);
\gpcolor{color=gp lt color 0}
\gpsetlinetype{gp lt plot 0}
\draw[gp path] (1.688,1.344)--(1.811,1.433)--(2.106,1.604)--(2.377,1.770)--(2.919,2.131)%
  --(3.435,2.547)--(4.100,3.111)--(4.912,3.869);
\gpcolor{color=gp lt color 1}
\gpsetpointsize{4.00}
\gppoint{gp mark 2}{(1.688,1.344)}
\gppoint{gp mark 2}{(1.811,1.433)}
\gppoint{gp mark 2}{(2.106,1.604)}
\gppoint{gp mark 2}{(2.377,1.770)}
\gppoint{gp mark 2}{(2.919,2.131)}
\gppoint{gp mark 2}{(3.435,2.547)}
\gppoint{gp mark 2}{(4.100,3.111)}
\gppoint{gp mark 2}{(4.912,3.869)}
\gpcolor{color=gp lt color border}
\gpsetlinetype{gp lt border}
\draw[gp path] (1.688,4.203)--(1.688,0.985)--(5.035,0.985)--(5.035,4.203)--cycle;
\gpdefrectangularnode{gp plot 1}{\pgfpoint{1.688cm}{0.985cm}}{\pgfpoint{5.035cm}{4.203cm}}
\end{tikzpicture}
}
		\caption{Bibliographical statistics on ML-ABM computed from author's bibliographic database and google scholar data on the Nov. 15, 2013}
		\label{pubperyear}
	\end{center}
\end{figure}

During last decade, ML-ABM has received significant and dramatically increasing interest~(fig.~\ref{pubperyear}). In this article we present a comprehensive and structured review of literature on the subject\footnote{The bibliographic database is available at the following URL: \url{http://www.lgi2a.univ-artois.fr/~morvan/Gildas_Morvan/ML-ABM_files/mlbib.bib}.}.

Another survey on the subject has been previously conducted by~\citeauthor{Gil-Quijano:2012} and published in different versions\footnote{This work was first published in french~\citep{Gil-Quijano:2009}, then extended~\citep{Gil-Quijano:2010},\citep[p. 185--204]{Louail:2010} and finally translated into english~\citep{Gil-Quijano:2012}}. While the present article aims at providing an overview of the literature,~\citeauthor{Gil-Quijano:2012} performed a comparative study of three models~\citep{Gil-Quijano:2008,Lepagnot:2009} and \citep{Pumain:2009a}. A similar survey, comparing four models, can be found in~\citet[p. 28--34]{Vo:2012b}. 

The paper is structured as follows. Section~\ref{Theoreticalissues} introduces the main theoretical contributions and section~\ref{Applicationdomains} presents the different application domains of ML-ABM, with an emphasis on social, flow, biological and biomedical models. In section~\ref{Discussion}, some issues about ML-ABM are discussed. Finally, the remaining sections conclude this paper by an analysis of the benefits, drawbacks and current limitations of the existing approaches.

\section{Theoretical issues}
\label{Theoreticalissues}

In the surveyed literature, three main theoretical issues have been addressed so far: 
\begin{itemize}
	\item the definition and implementation of meta-models and simulation engines,
	\item the detection and reification of emergent phenomena,
	\item and the definition of generic representations for aggregated entities.
\end{itemize}

They are described in the following sections.

\subsection{Meta-models, simulation engines and platforms}
\label{metamodels}

Many meta-models and simulation engines dedicated to ML-ABM have been proposed in the literature. They are are briefly presented in the following, in a chronological order.

Approaches based on DEVS have also been included~\citep{Zeigler:2000}. Indeed, DEVS, as a generic event-based simulation framework, has  been extended to support ABM~\citep{Duboz:2003,Duboz:2004,Muller:2009}. A comprehensive survey of the literature on multi-level DEVS extensions can be found in~\citet{Duhail:2013}. 

\textbf{GEAMAS}~\citep{Marcenac:1998b,Marcenac:1998,Marcenac:1998a} (GEneric Architecture for MultiAgent Simulation) is a pioneering ML-ABM framework integrating three levels of description (micro, meso, macro). Micro and macro levels represent respectively agent and system points of view while the meso (or middle) level represents an aggregation of agents in a specific context. Communication between levels is asynchronous. \textbf{GEAMAS-NG} is a newer version of the framework providing tools to detect and reify emergent phenomena~\citep{David:2011}.

\textbf{tMans}\footnote{\url{http://tmans.sourceforge.net/}}~\citep{Scheutz:2005} is a multi-scale agent-based meta-model and platform. Unfortunately, the project seems to have died in the bud.

\textbf{ML-DEVS}~\citep{Uhrmacher:2007,Steiniger:2012} is an extension of DEVS that allows the simulation of multi-scale models (and not only coupled models in which the behavior of a model is determined by the behaviors of its sub-models). Two types of relation between levels are defined: \textit{information propagation} and \textit{event activation}. However, ML-DEVS focuses on multi-scale modeling and therefore, only supports  pure hierarchies of models: \textit{interaction graphs} are viewed as \textit{trees}~\citep{Maus:2008}.

\textbf{CRIO}~\citep{Gaud:2008,Gaud:2008a,Gaud:2007} (Capacity Role Interaction Organization) is an organizational meta-model dedicated to ML-ABM based on the concept of holon~\citep{Koestler:1967,Koestler:1978}. It has been used to develop multi-scale simulations of pedestrian flows (cf. section~\ref{Flow}).


\textbf{SPARK}\footnote{\url{http://www.pitt.edu/~cirm/spark/}}~\citep{Solovyev:2010} (Simple Platform for Agent-based Representation of Knowledge) is a framework for multi-scale ABM, dedicated to biomedical research.

\textbf{~\citeauthor{Scerri:2010}} proposed a technical architecture to distribute the simulation of complex agent-based models~\citep{Scerri:2010}. This approach aims at integrating multiple ABMs in a single simulation, each ABM representing a specific aspect of the problem. In their article, authors focus on the management of time in such simulations. Therefore, the proposed platform provides two main technical services that ensure the consistency of simulations: (1) a \textit{time manager} that ensures that integrated ABMs advance in time in a consistent way and (2) a \textit{conflict resolver} that manages the problematic interactions between agents and shared data (such as the environment). Authors evaluate their approach on a modified version of Repast\footnote{\url{http://repast.sourceforge.net}} and show it can scale large-scale models easily.

\textbf{IRM4MLS}\footnote{\url{http://www.lgi2a.univ-artois.fr/~morvan/Gildas_Morvan/IRM4MLS.html}}~\citep{Morvan:2011,Morvan:2012b}  (Influence Reaction Model for Multi-Level Simulation) is a multi-level extension of IRM4S (Influence Reaction Model for Simulation)~\citep{Michel:2007}, an ABM meta-model based on  the Influence Reaction model which views action as a two step process: (1) agents produce "influences", \textit{i.e.}, individual decisions, according to their internal state and perceptions, (2) the system "reacts", \textit{i.e.}, computes the consequences of influences, according to the state of the world~\citep{Ferber:1996}. The relations of perception and influence between levels are specified with digraphs. IRM4MLS relies on a generic vision of multi-level modeling (see section~\ref{Introduction}). Therefore, interactions between levels are not constrained. It has been applied to simulate and control intelligent transportation systems composed of autonomous intelligent vehicles~\citep{Morvan:2012a,Morvan:2009a,Soyez:2011,Soyez:2013} (see section~\ref{Flow}).

\textbf{ML-Rules}~\citep{Maus:2011} is a rule-based multi-scale modeling language dedicated to cell biological systems. Rules, describing system dynamics, are described in a similar way as in chemical reaction equations. ML-Rules has been implemented within the simulation framework JAMES II\footnote{\url{http://www.jamesii.org}}. This approach does not refer explicitly to ABM; however, multi-level rule-based languages seem a promising way to engineer complex individual-based models.

\textbf{\citeauthor{Muller:2011a}} developed an approach that consists in decomposing a problem according to the complementary points of view involved in the modeling~\citep{Muller:2011,Muller:2011a,Muller:2012}. For instance in their case study the problem is the relation between residential and scholar segregation. Three points of view are considered: the geographer, the sociologist and the economist. Then, independent conceptual agent-based models are defined for each point of view. As models share agents and concepts, the conceptual models cannot be merged without some processing. Indeed, a same concept can have different meanings according to the point of view. To solve this issue, ~\citeauthor{Muller:2011a} adapt a technique described in the modular ontology literature: defining bridge rules that explicit the relations between concepts.

\textbf{PADAWAN}~\citep{Picault:2011} (Pattern for Accurate Design of Agent Worlds in Agent Nests) is a multi-scale ABM meta-model based on a compact matricial representation of interactions, leading to a simple and elegant simulation framework. This representation is based on the meta-model of IODA (Interaction-Oriented Design of Agent simulations) dedicated to classical (1-level) ABM~\citep{Kubera:2008}.

\textbf{GAMA}\footnote{\url{http://code.google.com/p/gama-platform/}}~\citep{Taillandier:2010, Taillandier:2012,Drogoul:2013} is an ABM platform with a dedicated modeling language, GAML, that offers multi-level capabilities. Moreover, it includes a framework (a set of predefined GAML commands) to \textit{agentify} emerging structures~\citep{Vo:2012}. It is certainly the most advanced platform, from an end-user point of view, that integrates  a multi-level approach. The multi-scale meta-model focuses on the notion of \textit{situated agent} and therefore, top class abstractions include geometry and topology of simulated entities~\citep{Vo:2012a}. The notion of level does not appear explicitly but the concept of \textit{species} defines attributes and behaviors of a class of same type agents and the multi-scale structure of the model, \textit{i.e.}, how species can be nested within each other.

\begin{figure}[t]
\begin{center}
\includegraphics[width=12cm]{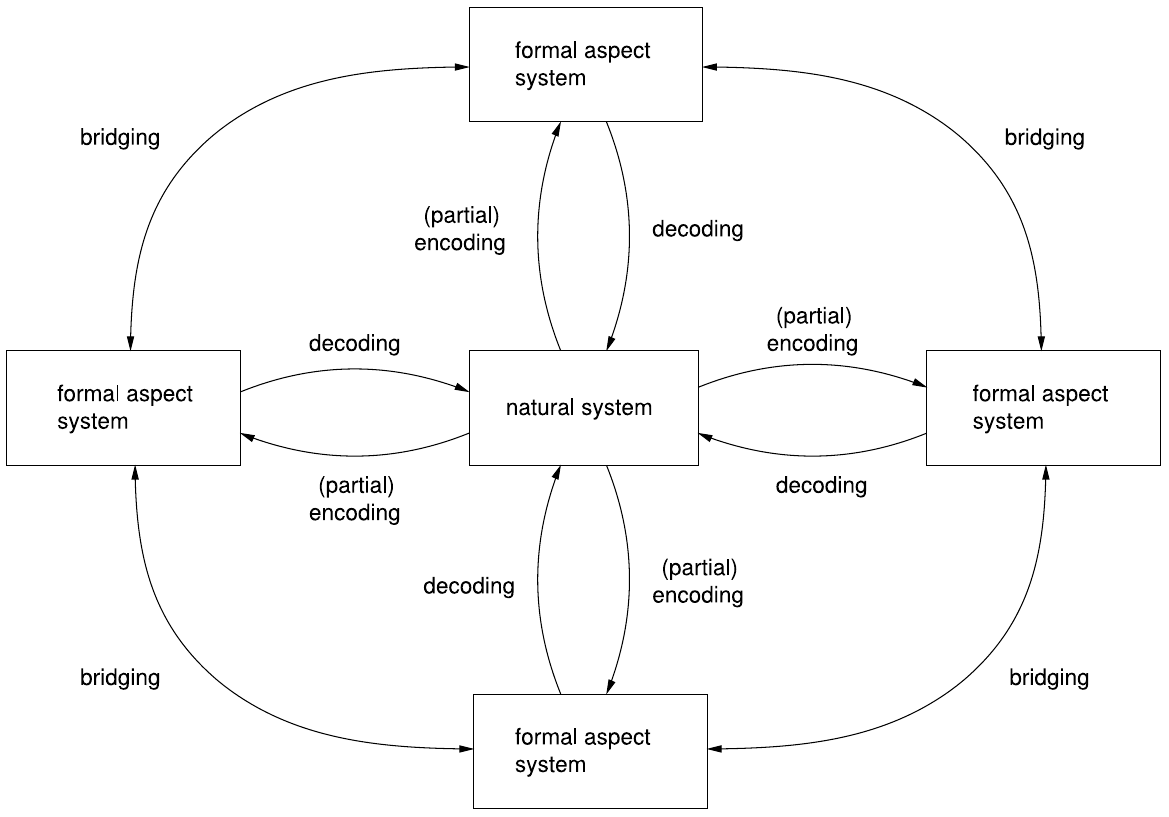}
\caption{The~\citet{Seck:2012} approach}
\label{seck2012}
\end{center}
\end{figure}

\textbf{\citeauthor{Seck:2012}} developed an extension of DEVS that allows the simulation of multi-level (\textit{i.e.}, non hierarchically coupled) models~\citep{Seck:2012}.  The coupling between levels is done through regular DEVS models, named bridge models~(fig.~\ref{seck2012}).

\textbf{AA4MM}~\citep{Camus:2012,Siebert:2010,Siebert:2011,Camus:2013} (Agent and Artifact for Multi-Modeling) is a multi-modeling (or model coupling) meta-model applied to ML-ABM. Levels are reified by agents that interact trough artifacts. This meta-model extends existing ones, see \textit{e.g.},~\citet{Bonneaud:2007,Bonneaud:2008}, distributing the scheduling between levels.

\subsection{Observation, detection and agentification of emergent phenomena}
\label{detection}

An important issue concerning ML-ABM is to observe, detect and possibly reify (or more precisely agentity) phenomena emerging from agent interactions. Of course, the question is not to detect any emergent phenomenon but those of interest, in order to \textit{e.g.}, adapt the level of detail of simulations, model cross-level interactions or observe multi-level behaviors.

Very different approaches have been proposed to solve this problem. The first ones were of course exploratory. Therefore, they rely on dedicated methods related to specific models. Newer works focus on generic methodologies and frameworks. They are briefly presented in a chronological order.  

\subsubsection{Dedicated clustering methods}
The pioneering RIVAGE project~\citep{Servat:1998,Servat:1998a,Servat:2000} aimed "at modeling runoff, erosion and infiltration on heterogeneous soil surfaces"~\citep[p. 184]{Servat:1998}. At the microscopic level, water is viewed as a set of interacting \textit{waterballs}. An indicator characterizes waterball movements to detect two types of remarkable situations: straight trajectories (corresponding to the formation of ravines) and stationary particles (corresponding to the formation of ponds). Close agents sharing such properties are aggregated in ravine or pond macroscopic agents.

\citet{Bertelle:2002,Tranouez:2001,Tranouez:2003,Tranouez:2005,Tranouez:2006, Tranouez:2009} aimed at changing the level of detail of fluid flow simulations using the vortex method~\citep{Leonard:1980}. The goal was, as in the RIVAGE model, to detect complex structures, \textit{i.e.}, clusters of particles sharing common properties, and aggregate them. However, the detection of emergent phenomena relies on graph-based clustering methods. \citet{Moncion:2010} used a similar approach to detect aggregations of agents in flocking simulations.

\citet{Gil-Quijano:2007,Gil-Quijano:2007a} and \citet{Gil-Quijano:2008} used various clustering  algorithms such as self-organizing maps, K-Means and particle swarm algorithms, to detect group formations. 

\subsubsection{Generic frameworks}

\citet{Chen:2008,Chen:2008a,Chen:2009,Chen:2009a,Chen:2010,Chen:2010a,Chen:2013} proposed a formalism, named \textit{complex event types} to describe multi-level behaviors in ABMs. "Conceptually, complex events are a configuration of simple events where each component event can be located in a region or point in a hyperspace that includes time, physical space and any other dimensions"~\citep[p. 4]{Chen:2008}. Using this approach, it is then possible to formally define and observe complex phenomena at different levels.
 
\citet{David:2008,David:2009} developed a conceptual and technical framework to handle emergence reification.
It is implemented in the GEAMAS-NG platform (cf. GEAMAS paragraph in the previous section) and has been used in a population model of the Reunion Island, to detect and reify new urban areas~\citep{David:2011,David:2012}. 

A similar framework has been integrated in the GAMA platform (see GAMA paragraph in the previous section). It includes various clustering methods developed in the literature~\citep{Vo:2012}.

SimAnalyzer is a general-purpose tool, to detect and describe group dynamics in simulations~\citep{Caillou:2012,Caillou:2012b}.

%
%
%

\subsection{Representation of aggregated entities}

While developing generic representations for aggregated entities (also called group abstractions) seems an important issue, to the best of our knowledge a few publications are available on the subject. The common idea in these works is to delegate the computation of agent behavioral functions to less detailed agents, representing groups of aggregated agents, in other  levels in order to reduce the complexity of interaction computing.

\citet{Sharpanskykh:2011,Sharpanskykh:2011a} proposed two approaches to group abstraction dedicated to models where agent state are represented by variables taking values in $\{0, 1\}$ or $[0, 1]$:
\begin{itemize}
	\item \textit{weighting averaging}: an aggregated state of a group is estimated by averaging  agent states, with a weighting factor related to the strength of influence of an agent in the group (the stronger, the more important).
	\item \textit{invariant-based abstraction}: this approach consists in determining an invariant in a group of agents, \textit{i.e.}, a property that does not change in time and using it as a conservation law.
\end{itemize}
\citeauthor{Sharpanskykh:2011} applied these methods to a collective decision making model of social diffusion. They performed a comparative study of the methods, focusing on computational efficiency and approximation error.

\citet{Parunak:2012} introduced the notion of \textit{pheromone field} (refering to the concept of \textit{mean field} in statistical physics) that "gives the probability of encountering an agent of the type represented by the field at a given location"~\citep[p. 115]{Parunak:2012}. In this approach, agents act according to their perceptions of pheromone fields (but not of agents).

\citet{Navarro:2012} proposed a generic approach based on the notion of \textit{mesoscopic representation}: agents sharing common properties (related to their physical or mental states) delegate the computation of behavioral functions  to a mesoscopic agent. Authors developed this approach to reduce the computational cost of simulations while guaranteeing accurate results.

\section{Application domains}
\label{Applicationdomains}

ML-ABM has been used in various fields such as
\begin{itemize}
	\item biomedical research
	\begin{itemize}
		\item cancer modeling~\citep{Andasari:2012,Brown:2011,Deisboeck:2010,Lepagnot:2009,Olsen:2013,Paiva:2009,Rejniak:2011,Sun:2012,Wang:2007,Wang:2008,Wang:2008a,Wang:2013,Zhang:2007,Zhang:2009a,Zhang:2011,Zhang:2009},
		\item inflammation modeling~\citep{An:2008,An:2009,An:2013,Kim:2012,Scheff:2012,Vodovotz:2008,Wakeland:2007},
		\item arterial adaptation~\citep{Hayenga:2011,Thorne:2011}, 
		\item stent design~\citep{Tahir:2011},
		\item vascular tissue engineering~\citep{Zahedmanesh:2012},
		\item  bone remodeling~\citep{Cacciagrano:2010},
	\end{itemize}
	\item   flow modeling of walking (and running)~\citep{Gaud:2008,Navarro:2011,Nguyen:2011,Nguyen:2012,Xi:2012}, driving~\citep{Espie:2006,Mammar:2006,Bourrel:2002,Bourrel:2003,Bourrel:2003a,Magne:2000,Poschinger:2002,Burghout:2005,El-hmam:2006,El-hmam:2006a,El-Hmam:2006b,El-hmam:2008,Morvan:2012a,Morvan:2009a,Sewall:2011,Soyez:2011,Soyez:2013,Wedde:2012} or streaming~\citep{Servat:1998,Servat:1998a,Tranouez:2006} agents,
	\item biology~\citep{Adra:2010,Biggs:2013,Christley:2007,Christley:2007a,Jeschke:2008,Marino:2011,Montagna:2010,Montagna:2010a,Seal:2011,Shimoni:2011,Smallwood:2006,Smallwood:2010,Stiegelmeyer:2013,Sun:2009},
	\item  social simulation~\citep{Conte:1996,Conte:2007,Dascalu:2009,Dascalu:2011,Gil-Quijano:2008,Hassoumi:2012,Laperriere:2012,Louail:2010,North:2010,Ozik:2008,Parry:2012,Pumain:2009a,Sawyer:2001,Sawyer:2003,Schaller:2012,Schillo:2001,Seck:2012,Squazzoni:2008},
	\item ecology~\citep{Belem:2009,Belem:2009a,Belem:2013,Cheong:2012,Duboz:2003,Duboz:2004,Le:2011,Marilleau:2008,Morvan:2008,Morvan:2009a,Prevost:2004,Ratze:2007,Rounsevell:2012,Seidl:2012,Seidl:2010,Semeniuk:2011,Schmidt:2011,Vincenot:2011},
	\item military simulation~\citep{Parunak:2009,Mathieu:2007,Mathieu:2007a}.
\end{itemize} 

An interesting comparative analysis of three of these models can be found in~\citet{Gil-Quijano:2009,Gil-Quijano:2010,Gil-Quijano:2012} and \citet[p. 185--204]{Louail:2010}.

\subsection{Social simulation} 

Social simulation is defined by~\citet[p. 4]{Squazzoni:2008} as "the study of social outcomes, let us say a macro regularity, by means of computer simulation where agents' behavior, interactions among agents and the environment are explicitly modeled to explore those micro-based assumptions that explain the macro regularity of interest".

Major social theories developed in the second half of the twentieth century, \textit{e.g.}, \textit{structuration}~\citep{Giddens:1987} and \textit{habitus}~\citep{Bourdieu:1994} theories\footnote{These theories are described by some sociologists as hybrid~\citep{Sawyer:2001}.}, share a common ambition: solving the micro/macro (so called \textit{agency/structure}) problem that can be summarized by the following question:  \textit{To understand social systems, should we observe agent interactions (micro level) or structures emerging from these interactions (macro level)?} Such theories tend to consider altogether agent positions in the social space (objective facts) and goals (subjective facts) to explain their beliefs and actions. Their answer to the previous question could be: \textit{social systems  can only be understood by considering simultaneously agent interactions and structures in which they occur}:
\begin{center}
\begin{tikzpicture}[>=stealth',shorten >=1pt,auto,semithick]
\node (l1) {social structures} ;
\node[below of=l1,node distance=1.5cm] (l2) {agent interactions.} ;
\draw[<->] (l1) -- node[right] {social practices} (l2)   ;
\end{tikzpicture} 
\end{center}

A key concept used by social theorists and modelers to understand downward (or top-down) causation in social systems, \textit{i.e.}, how social structures influence agents, is \textit{reflexivity}. It can be defined as the "regular exercise of the mental ability, shared by all normal people, to consider themselves in relation to their (social) contexts and vice versa"~\citep[p. 4]{Archer:2007}. Thus, social systems differ from other types of systems, by the reflexive control that agents have on their actions: "The reflexive capacities of the human actor are characteristically involved in a continuous manner with the flow of day-to-day conduct in the contexts of social activity"~\citep[p. 22]{Giddens:1987}. Two very different approaches, both from technical and methodological perspectives, can be considered to simulate systems composed of reflexive agents:
\begin{itemize}
\item a purely emergentist approach, only based on the cognitive capabilities of agents to represent and consider themselves in relation to the structures emerging from their interactions ---~\textit{e.g.},~\citet{Conte:2002,Gilbert:2002}, 
\item a multi-level approach based on the cognitive capabilities of agents and the dynamic \textit{reification} of interactions between social structures and agents, \textit{i.e.}, processes that underlie social practices ---~\textit{e.g.},~\citet{Gil-Quijano:2009,Pumain:2009}. 
\end{itemize}
According to~\citet{Giddens:1987}, two forms of reflexivity can be distinguished: \textit{practical}  (agents are not conscious of their reflexive capabilities, and therefore, are not able to resonate about them) and \textit{discursive} (agents are conscious of their reflexive capabilities)  reflexivity. These two forms are respectively related to the ideas of \textit{immergence} in which agent interactions produce emergent properties that modify the way they produce them~\citep{Conte:2007} and \textit{second order emergence} in which agent interactions produce emergent properties that are recognized (incorporated) by agents and influence their actions~\citep{Gilbert:2002}. 

\begin{figure}[h]
\begin{center}
\begin{tikzpicture}[>=stealth',shorten >=1pt,auto,semithick]
\node (micro) {\textit{microsociology}} ;
\node[above of=micro,node distance=1.5cm]  (meso) {\textit{mesosociology}} ;
\node[above of=meso,node distance=1.5cm]   (macro) {\textit{macrosociology}} ;

\node[right of=micro,node distance=5cm]  (agent) {agent} ;
\node[right of=meso,node distance=3cm]  (community) {community} ;
\node[right of=meso,node distance=7cm]  (organization) {organization} ;
\node[right of=macro,node distance=5cm]  (large) {large-scale social structures} ;
\draw[<->] (agent) -- node {} (community)   ;
\draw[<->] (agent) -- node {} (organization)   ;
\draw[<->] (organization) -- node {} (community)   ;
\draw[<->] (organization) -- node {} (large)   ;
\draw[<->] (community) -- node {} (large)   ;
\end{tikzpicture} 
\caption{ML-ABM in social simulation as a link between concepts defined at different levels}
\label{mlmsocio}
\end{center}
\end{figure}
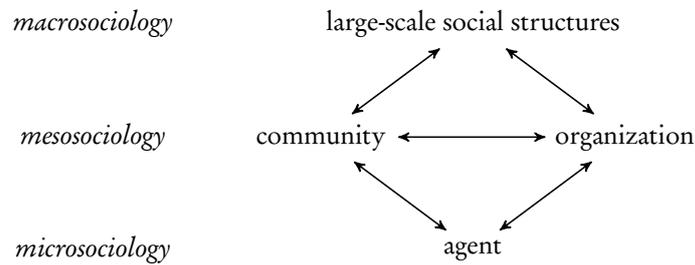

ML-ABM can also be viewed as a way to link independent social theories (and therefore concepts) defined at different levels~(fig.~\ref{mlmsocio})~\citep{Sawyer:2001,Sawyer:2003,Seck:2012}. Thus,~\citet{Seck:2012} proposed a model of social conflicts integrating agent behaviors and social laws. \citet{Gil-Quijano:2007,Gil-Quijano:2007b,Gil-Quijano:2008}  developed a multi-scale model of intra-urban mobility. The microscopic level represents households and housing-units, the mesoscopic one, groups of micro-agents and urban-sectors and the macroscopic one, the city itself. A clustering algorithm is applied to detect and reify groups of households and housing-units. \citet{Pumain:2009a} developed \textit{Simpop3}, a multi-scale model based on two previously developed single-scale models: \textit{Simpop nano}, simulating the internal dynamics of a city and \textit{Simpop2}, simulating city interactions. 

Readers interested in a more comprehensive presentation of these questions may refer to~\citet{Schillo:2001,Sawyer:2003,Squazzoni:2008} and \citet{Raub:2011} .

\subsection{Flow modeling}
\label{Flow}

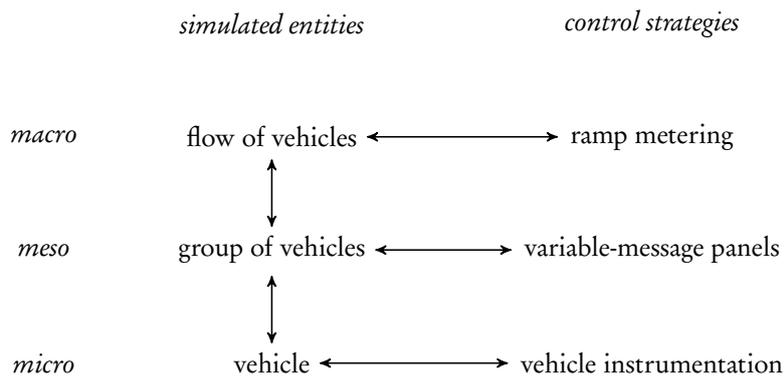
\begin{figure}[h]
\begin{center}
\begin{tikzpicture}[>=stealth',shorten >=1pt,auto,semithick]
\node (micro) {\textit{micro}} ;
\node[above of=micro,node distance=1.5cm]  (meso) {\textit{meso}} ;
\node[above of=meso,node distance=1.5cm]   (macro) {\textit{macro}} ;

\node[right of=micro,node distance=3cm]  (vehicle) {vehicle} ;
\node[right of=meso,node distance=3cm]  (group) {group of vehicles} ;
\node[right of=macro,node distance=3cm]  (flow) {flow of vehicles} ;
\node[right of=micro,node distance=8cm]  (instrumentation) {vehicle instrumentation} ;
\node[right of=meso,node distance=8cm]  (panels) {variable-message panels} ;
\node[right of=macro,node distance=8cm]  (metering) {ramp metering} ;

\node[above of=flow,node distance=1.5cm] {\textit{simulated entities}} ;
\node[above of=metering,node distance=1.5cm] {\textit{control strategies}} ;

\draw[<->] (vehicle) -- node {} (group)   ;
\draw[<->] (group) -- node {} (flow)   ;
\draw[<->] (vehicle) -- node {} (instrumentation)   ;
\draw[<->] (group) -- node {} (panels)   ;
\draw[<->] (flow) -- node {} (metering)   ;
\end{tikzpicture} 
\caption{ML-ABM and control in traffic simulation}
\label{mlmtraffic}
\end{center}
\end{figure}

A flow of moving agents can be observed at different scales. Thus, in traffic modeling, three levels are generally considered: the \textit{micro}, \textit{meso} and \textit{macro} levels, representing respectively the interactions between vehicles, groups of vehicles sharing common properties (such as a common destination or a common localisation) and flows of vehicles. Each approach is useful in a given context: micro and meso models allow to simulate road networks with complex topologies such as urban area, while macro models allow to develop control strategies to prevent congestion in highways. However, to simulate large-scale road networks, it can be interesting to integrate these different representations~(fig.~\ref{mlmtraffic}).  The main problem is to determine an appropriate coupling between the different representations, \textit{i.e.}, that preserves the consistency of simulations~\citep{Davis:1993}.

\subsubsection{Micro-macro models}

An interesting comparison of existing micro-macro traffic models can be found in~\citet[p. 31--44]{El-hmam:2006c}~(table~\ref{hybridmodels}). The coupling methods rely on the following idea: creating a virtual section which is both macro and micro at level connections. This virtual section is used as a buffer to generate appropriate micro or macro data according to the type of connection. On this basis, \citet{El-hmam:2006,El-hmam:2006a,El-Hmam:2006b,El-hmam:2006c,El-hmam:2008} proposed a generic coupling method between agent-based microscopic models and widely used macroscopic models such as LWR, ARZ and Payne.

While micro-macro flow models were essentially developed in the traffic domain, other applications such as crowd simulation emerged in recent years~\citep{Nguyen:2011,Nguyen:2012}.

However, all the surveyed hybrid models share the same limitation: connections between levels are fixed \textit{a priori} and cannot be changed at runtime. Therefore, to be able to observe some emerging phenomena such as congestion formation or to find the exact location of a jam in a large macro section, a dynamic hybrid modeling approach is needed~\citep{Sewall:2011}.

\begin{table}[h]
\begin{center}
\begin{tabular}{ccc}
\hline
model&micro model&macro model\\
\hline
\citet{Magne:2000}&SITRA-B+&SIMRES\\
\citet{Poschinger:2002}&IDM&Payne\\
\citet{Bourrel:2003}&\multirow{2}{*}{optimal velocity}&LWR\\
\citet{Mammar:2006}&&ARZ\\
\citet{Espie:2006}&ARCHISM&SSMT\\
\citet{El-hmam:2006c}&\multirow{2}{*}{generic ABM}&LWR, ARZ, Payne\\
\citet{Sewall:2011}&&ARZ\\
\hline
\end{tabular}
\caption{Main micro-macro traffic flow models, adapted from~\citet[p. 42]{El-hmam:2006c}} \label{hybridmodels}
\end{center}
\end{table}%
 


\subsubsection{Micro-meso models}

This kind of models is often used to reduce the complexity of agent interactions.
Agents sharing common properties can be aggregated to form up a higher level (mesoscopic) agent and then, save computer resources or describe group dynamics such as in the already mentioned RIVAGE~\citep{Servat:1998,Servat:1998a,Servat:2000} and DS~\citep{David:2011} models~(cf. section~\ref{detection}). Conversely, mesoscopic agents can be broken up into lower level agents if related structures vanish.

\citet{Morvan:2012a,Morvan:2009a} introduced an multi-level approach to solve the dead-lock problem in field-driven autonomous intelligent vehicle systems. These systems generally rely on self-organization to achieve their goals, but AIVs can remain trapped into dead-locks. When such a situation is detected (using a similar approach than~\citet{Servat:1998,Servat:1998a,Servat:2000}), it is agentified to solve the problem using hierarchical control.

\citet{Navarro:2011,Flacher:2012,Navarro:2012,Navarro:2013} proposed an innovative framework for such models: (de)aggregation functions rely not only on the observable state of simulations (the environment) but also on the internal state of agents. It has been applied to pedestrian flow simulation. The proximity between agent states (external and internal) is computed by an affinity function. 

\begin{figure}[t]
\begin{center}
\begin{tikzpicture}

\node(a)[draw] at (0,0) [rounded corners]{
            \begin{tabular}{c}
	spiritAgent
\end{tabular}};

\node(b)[draw] at (2,2)[rounded corners]{
            \begin{tabular}{c}
	conceptualAgent
\end{tabular}};

\node(c)[draw] at (4,0)[rounded corners]{
            \begin{tabular}{c}
	bodyAgent
\end{tabular}};

\node(d)[draw] at (8,0)[rounded corners]{
            \begin{tabular}{c}
	level
\end{tabular}};

\node(e)[draw] at (6,2)[rounded corners]{
            \begin{tabular}{c}
	environment
\end{tabular}};

\draw (e.south) -- (d.north) 
	node[very near end, left] {0..n}
	node[near start, left]{1};

\draw (c.north) -- (e.south)
	node[very near start, right] {0..n}
	node[near end, right]{1};

\draw (c.east) -- (d.west)
	node[near start, below] {0..n}
	node[very near end, below]{1};

\draw (a.east) -- (c.west)
	node[very near start, below]{1}
	node[near end,below]{1..n};

\draw [-diamond] (c.north) -- (b.south)
	node[very near start, left]{1..n}
	node[near end, below]{1};

\draw [-diamond] (a.north) -- (b.south)
	node[very near start, right]{1}
	node[near end, below]{1};
\end{tikzpicture}
\caption{Mind/bodies separation in the~\citet{Soyez:2013} model}
\label{soyezMindBody}
\end{center}
\end{figure}
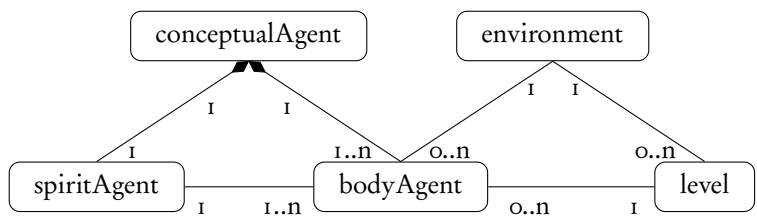

\citet{Soyez:2011,Soyez:2013} extended this framework on the basis of IRM4MLS. Agents are "cut" into a a set of physical parts (bodies), situated in different levels, and a non-situated part (mind) (see fig.~\ref{soyezMindBody}). Therefore, these different parts can be (de)aggregated independently. This approach has been applied to dynamically adapt the level of detail in a port operations simulator.

\subsection{Biological and biomedical models}

A biological system can be considered at different levels of organization:
\begin{center}
... $\rightarrow$ molecule $\rightarrow$  cell $\rightarrow$ tissue  $\rightarrow$ organ  $\rightarrow$ ... , 
\end{center}
that basically correspond to the segmentation of biological research into specialized communities:
\begin{center}
... $\rightarrow$ molecular biology $\rightarrow$ cell biology $\rightarrow$ histology $\rightarrow$ physiology  $\rightarrow$ ... .
\end{center}

Each research area has developed its own ontologies and models to describe the same reality observed at different levels. However, this reductionist approach fails when addressing complex issues~\citep{Schnell:2007}. Thus, it has been shown that living systems are co-produced by processes at different levels of organization~\citep{Maturana:1980}. Therefore, an explanatory model of such systems should account for the interactions between levels. 

\subsubsection{Cell biology}

At least two levels are explicitly represented in cell biology models: the \textit{macroscopic} one, representing the extracellular environment and the interactions between cells, and the \textit{microscopic} one, representing the intracellular environment and the interactions between cell components such as signaling pathways and gene regulatory networks. A major modeling issue is that these two levels continuously influence each other. It leaded to the development of multi-scale models.

\citet{Montagna:2010,Montagna:2010a} developed a model of morphogenesis in biological systems, in particular for the \textit{Drosophila Melanogaster} species. \citet{Maus:2011} proposed a model of \textit{Schizosaccharomyces pombe} (a species of yeast) cell division and mating type switching based on the ML-Rules approach (cf. section~\ref{metamodels}).

\subsubsection{Cancer modeling}

Cancer is a complex spatialized multi-scale process, starting from genetic mutations and potentially leading to metastasis. Moreover, it has multi-scale (from from genetic to environmental) causes.  Therefore, it can be studied from  various perspectives from the intracellular (molecular) to the population levels.

ML-ABM is a promising paradigm to model cancer development~\citep{Wang:2008}. Indeed, as~\citet[p. 140]{Schnell:2007} note, "a multi-scale model would allow us to explore the effect of combination therapies, approaches that attempt to stop cancer in its tracks by barricading multiple pathways. Most present models, focusing on processes at a single scale, cannot provide this comprehensive view."

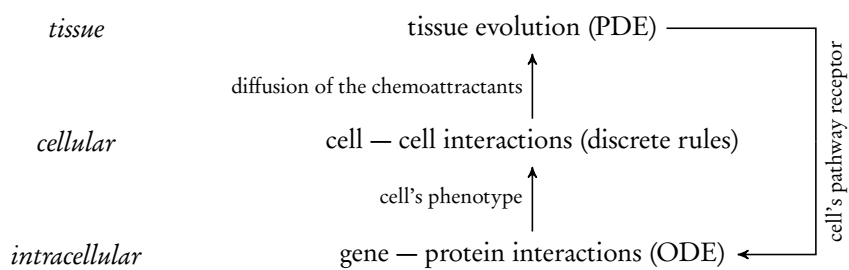
\begin{figure}[ht]
\begin{center}
\begin{tikzpicture}[>=stealth',shorten >=1pt,auto,semithick]
\node (micro) {\textit{intracellular}} ;
\node[above of=micro,node distance=1.5cm]  (meso) {\textit{cellular}} ;
\node[above of=meso,node distance=1.5cm]   (macro) {\textit{tissue}} ;

\node[right of=micro,node distance=6cm]  (vehicle) {gene --- protein interactions (ODE)} ;
\node[right of=meso,node distance=6cm]  (group) {cell --- cell interactions (discrete rules)} ;
\node[right of=macro,node distance=6cm]  (flow) {tissue evolution (PDE)} ;

\node[right of=group,node distance=4cm,rotate=90]  (nada) {\footnotesize{cell's pathway receptor}} ;

\path (vehicle) edge[->] node {\footnotesize{cell's phenotype}} (group)   ;
\path (group) edge[->] node {\footnotesize{diffusion of the chemoattractants}} (flow)   ;

\path [draw] (flow) -| (nada.north)+(0,-0.2cm);
\path [draw,->] (nada.north) |- (vehicle);

\end{tikzpicture} 
\caption{ML-ABM in brain tumor modeling~\citep{Zhang:2007,Zhang:2009,Zhang:2009a,Zhang:2011}}
\label{mlmcancer}
\end{center}
\end{figure}

\citeauthor{Zhang:2007} developed an ML-ABM of a brain tumor named \textit{Glioblastoma Multiforme} (GBM)~\citep{Zhang:2007,Zhang:2009,Zhang:2009a,Zhang:2011}. This model explicitly defines the relations between scales and uses different modeling approaches: ordinary differential equations (ODE) at the intracellular level, discrete rules typically found in ABM at the cellular level and partial differential equations (PDE) at the tissue level (see fig.~\ref{mlmcancer}).

Moreover, this model also relies on a multi-resolution approach: heterogenous clusters, \textit{i.e.}, composed of migrating and proliferating cells are simulated at a high resolution while homogenous clusters of dead cells are simulated at a lower resolution. In short, "more computational resource is allocated to heterogenous regions of the cancer and less to homogenous regions"~\citep[p. 6]{Zhang:2011}. This model has been implemented on graphics  processing units (GPU), leading to an efficient parallel simulator~\citep{Zhang:2011}.

\citet{Sun:2012} also developed a brain tumor ML-ABM available as a MATLAB library called ABM-TKI\footnote{\url{https://sites.google.com/site/agentbasedtumormodeling/home}}. It is based on a 4 level architecture (tissue, microenvironmental, cellular, modelcular). 

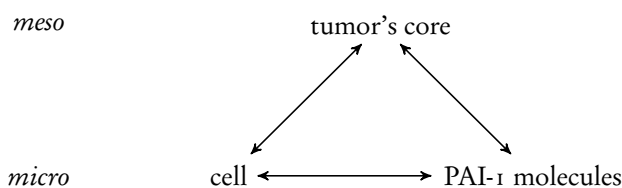
\begin{figure}[ht]
\begin{center}
\begin{tikzpicture}[>=stealth',shorten >=1pt,auto,semithick]

\node (micro) {\textit{micro}} ;
\node[above of=micro,node distance=2cm]  (meso) {\textit{meso}} ;

\node[right of=micro,node distance=2.5cm]  (cell) {cell} ;
\node[right of=micro,node distance=6.5cm]  (pai) {PAI-1 molecules} ;

\node[right of=meso,node distance=4.5cm]  (core) {tumor's core} ;

\path (cell) edge[<->] node {} (pai)
	 (core) edge[<->] node {} (pai)  
	 (core) edge[<->] node {} (cell)   ;

\end{tikzpicture} 
\caption{ML-ABM in avascular tumor growth modeling~\citep{Lepagnot:2009}}
\label{mlmcancer2}
\end{center}
\end{figure}

\citet{Lepagnot:2009} model the growth of avascular tumors to study the impact of PAI-1 molecules on metastasis. To deal with the problem complexity (a tumor may be composed of millions of cells) two levels are introduced: the cell and the tumor's core levels (fig.~\ref{mlmcancer2}). Indeed, such cancers are generally structured as a kernel of necrosed or quiescent cells surrounded by living tumor cells. As necrosed and quiescent cells are mostly inactive, tumor's core is reified as a single upper-level agent, interacting with cells and PAI-1 molecules at its boundary. A more comprehensive analysis of this model can be found in~\citet{Gil-Quijano:2012}.

\subsection{Ecology}

Ecologists study processes that can have very different spatio-temporal dynamics. Then, characterizing their interactions is a complicated problem and traditional bottom-up or top-down approaches do not seem relevant: ABMs tend to be too complex, requiring a lot of computational resources\footnote{An interesting solution to this problem is to reduce the complexity of agent interactions using estimation algorithms such as the fast multipole method~\citep{Razavi:2011}.} while EBMs cannot deal with complex heterogenous environments~\citep{Shnerb:2000}.

Ecological systems are generally described as hierarchies~\citep{Muller:2005,Ratze:2007}. Thus the hierarchy theory is "a view of ecological systems, which takes the scales of observation explicitly into account and which tries to conceptualize the phenomena at their proper scale"~\citep[p. 14]{Ratze:2007}. ML-ABM seems a interesting way to implement this concept. Different modeling issues in Ecology have been solved by ML-ABM.

\begin{figure}[ht]
\begin{center}
\begin{tikzpicture}[>=stealth',shorten >=1pt,auto,semithick]

\node (abm) {ABM} ;

\node[right of=abm,node distance=4cm]  (ebm) {EBM} ;

\path (abm) edge[->, bend left] node {emergent computation} (ebm)
	 (ebm) edge[->,bend left] node {environment parametrization} (abm)   ;

\end{tikzpicture} 
\caption{The scale transfer approach~\citep{Duboz:2003,Duboz:2004}}
\label{mlmduboz}
\end{center}
\end{figure}
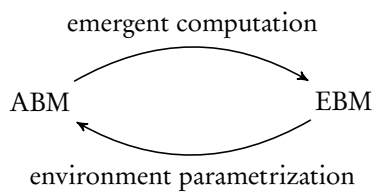

\citet{Duboz:2003,Duboz:2004} proposed the \textbf{scale transfer} approach to link microscopic and macroscopic models: the state of the system is computed by an ABM and is used to parametrize an EBM describing population dynamics. This EBM can then be used to parametrize the ABM environment (fig.~\ref{mlmduboz}).

\citet{Marilleau:2008} introduced a efficient method to represent complex soils,  named \textbf{APSF} (Agent, Pores, Solid and fractal). Traditionally, the environment is viewed as a regular grid, discretized into cells. A cell can represent a pore, \textit{i.e.}, a part of a soil cavity, a solid or a fractal. The idea is that a cell is not necessarily an atomic element describing an homogenous area but can be fractal, \textit{i.e.}, composed of smaller pore, solid or fractal cells with a self similar structure. Fractal cells are instantiated at run time, generating finer representations of the environment when it is needed. Thus, this approach based on self-generation allows to represent complex multi-scale environments at a minimal computational cost. It has been used in the SWORM (Simulated WORMS) model that studies the relation between earthworm activity and soil structure~\citep{Blanchart:2009,Laville:2012}.

\begin{figure}[ht]
\begin{center}
\begin{tikzpicture}[>=stealth',shorten >=1pt,auto,semithick]

\node (micro) {\textit{micro}} ;
\node[above of=micro,node distance=2cm]  (meso) {\textit{meso}} ;

\node[right of=micro,node distance=6cm]  (abm) {\textit{Diptera} larvae (ABM)} ;

\node[right of=meso,node distance=3cm]  (ca) {thermal dynamics (CA)} ;
\node[right of=meso,node distance=9cm]  (ebm) {maggot mass effect (EBM)} ;

\path (abm) edge[->] node {} (ebm)
	 (ca) edge[->] node {} (abm)  
	 (ebm) edge[->] node {} (ca)   ;

\end{tikzpicture} 
\caption{Levels of organization in a ML-ABM of necrophagous \textit{Diptera} development~\citep{Morvan:2008,Morvan:2009a}}
\label{mlmdiptera}
\end{center}
\end{figure}
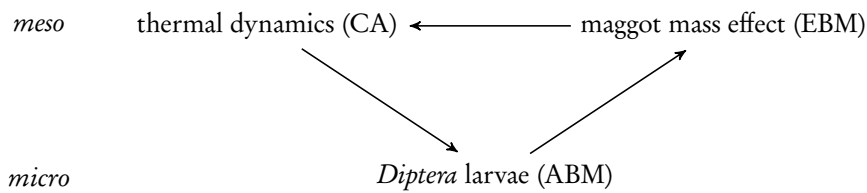

\textit{Diptera} larvae have a complex gregarious behavior that lead to the formation of large groups in which individuals regulate the temperature to optimize their development speed. This phenomenon can be described by a mesoscopic equation-based model (knowing the mass of the group and and the external temperature), while the crowding behavior of larvae can be modeled by an ABM. Moreover, the thermal dynamics of the cadaver can be modeled by a Cellular Automaton (CA). \citet{Morvan:2008,Morvan:2009a} integrated these different models in a ML-ABM to perform more accurate forensic entomology expertises. In this model, the EBM is parametrized according to the ABM state. It computes the increase of temperature caused by  \textit{Diptera} interactions at the group level and send it to the CA model that is used as an environment for the ABM~(fig.~\ref{mlmdiptera}). The environment can thus be viewed as an artifact, used to synchronize the different models.

\section{Discussion}
\label{Discussion}

In this section two issues are discussed:
\begin{itemize}
\item the different forms of level integration,
\item the use of ML-ABM to solve technical problems.
\end{itemize}

\subsection{Level integration}

In the introduction, ML-ABM has been defined as integrating heterogenous ABMs in a single model. Following the approach of \citet{Michel:2003} on interaction, one can distinguish at least two forms of integration:
\begin{itemize}
\item \textbf{weak integration}: levels share objects, \textit{e.g.}, environment properties, but not agents,
\item \textbf{strong integration}: levels share objects and agents. 
\end{itemize}

Weak integration can be regarded as a form of multi-modeling (or model coupling) where levels represent different models interacting through shared variables called artifacts~\citep{Camus:2012,Seck:2012,Siebert:2010}.

Weaker forms of integration are not regarded as multi-level modeling. Thus, in the SWARM platform~\citep{Minar:1996}, integration can be described as bottom-up or isotropic (information flows in one direction). An agent is designed as a russian doll and its behavior at a given level depends on the lower ones.

An simple example of strong integration is given by~\citet[p. 334]{Picault:2011}: "a membrane protein, which has an end inside the cell, and the other end outside".

In the meta-models presented in section~\ref{metamodels}, IRM4MLS and PADAWAN are, to the best of our knowledge, the only ones able to simulate strongly integrated levels. It is not surprising as they are based on formal interaction models (respectively  IRM4S and IODA) that differentiate between agent influences and level reaction and therefore, are able to represent \textbf{strong interaction}~\citep{Michel:2003}. In short, agents are strongly interacting if the interaction output depends on the influences of each agent. Thus, STRIPS-like action models (\textit{i.e.}, that view action as a change of the state of the world), used in most of agent-based simulation platforms, are unable to represent such interactions. Yet, \citep{Michel:2003}~showed that modeling a strong  interaction as a weak one leads to arbitrary implementation choices and result interpretation issues.

In ML-ABM, the problem is similar since levels can be seen are strongly interacting entities. Thus, we can conclude that a modeling formalism capable of representing strong interaction\footnote{Some authors described these approaches as "interaction-based modeling", by opposition to the term "individual-based modeling", as that they focus on interactions rather than on individual behaviors~\citep{Desmeulles:2009,Kubera:2008}.} can be extended to a multi-level one.


\subsection{Multi-level technical tools}

\begin{figure}[ht]
\begin{center}
\tikzstyle{blocdebase}= [draw, text centered,minimum height=0.5cm, rounded corners]
\begin{tikzpicture}[>=stealth',shorten >=1pt,auto,semithick]

\node (input) {\footnotesize{Inputs}} ;

\node[blocdebase,below of=input,node distance=1cm]  (simulation) {Simulation} ;
\node[blocdebase,below of=simulation,node distance=1cm]  (detection) {Detection} ;
\node[below of=detection,node distance=1cm]  (visualization) {\footnotesize{Visualization tools}} ;

\node[right of=simulation,node distance=2cm, label={[rotate=90,shift={(-0.6,-0.4)}]\footnotesize{Agentification}}]  (nada) {};

\path (input) edge[->] node {} (simulation)   ;
\path (simulation) edge[->] node[left] {\footnotesize{emergent phenomena}} (detection)   ;
\path (detection) edge[->] node {} (visualization)   ;

\path [draw] (detection) -| (nada.west);
\path [draw,->] (nada.west) |- (simulation);

\end{tikzpicture} 
\caption{Two main uses of detected emergent phenomena: visualization or re-injection in the simulation as agents}
\label{mlmvisualization}
\end{center}
\end{figure}
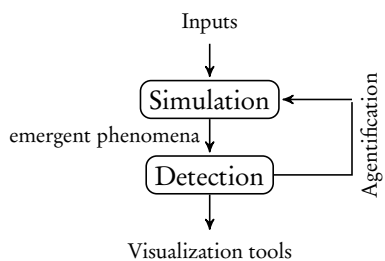

Although considering cross-level interactions is usually related to the application domain as shown previously in this article, it can also be viewed as a technical tool:
\begin{itemize}
\item an ABM (microscopic level) can be used to parametrize an equation based model (macroscopic level)~\citep{Duboz:2003,Duboz:2004,Nguyen:2012a},
\item levels can be created at run-time by other levels to generate fractal environments~\citep{Marilleau:2008},
\item a mesoscopic level can be viewed as a controller (in the control theory meaning) of group-related properties~\citep{Morvan:2012b}.
\item automated observation and analysis tools can be introduced at levels not explicitly present in the model to detect and study (multi-level) emergent phenomena\footnote{Indeed, as~\citet[p. 4]{An:2008} notes about ABM, "since the models rely on an ill-defined principle of 'emergence' in order to transcend the epistemological boundaries represented by the multiple hierarchies of system organization, their behavior is difficult to characterize analytically".} (cf. section~\ref{detection}). However, reified emergent phenomena cannot be considered as model entities since they are not re-injected in the simulation~(see fig.~\ref{mlmvisualization}),
\item a radical interpretation of ML-ABM is the concept of multi-future~\citep{Parunak:2010,Parunak:2010a}. The possible trajectories of agent actions are computed by "ghosts" as a pheromone field and agents act according to it, selecting the most probable one.
\end{itemize}

As these ideas are domain-independent, they could be implemented as a generic library, providing such services to a classical ABM or ML-ABM framework or platform.

%
%


\section{Conclusion}
\label{conclusion}

An important challenge for the scientific community is to find ways to deal with the complexity of natural and artificial complex systems. This issue led to the development of dedicated modeling paradigms and engineering principles that focus on interaction and organization. We strongly believe that such techniques will play an important role in the future.

In this article, we conducted a comprehensive analysis  of the bibliography available on one of them: multi-level agent-based modeling. 

As we shown, many papers focus on the application of this technique and then, are published in domain-related journals and conferences. However, a dedicated venue for theoretical or methodological papers is lacking. Such forum would allow to unify the vocabulary and concept definitions, discuss the main issues of this approach and more generally, share ideas with the interested communities.

\begin{multicols}{2}
\section*{Acknowledgments}

I am particularly grateful to my colleagues Shahin Gelareh, Yoann Kubera and Jean-Baptiste Soyez for their help and support. I also would like to thank Chih-Chun Chen, Jean-Pierre Müller and Sebastian Senge for suggesting new entries in the bibliographical database.

\addcontentsline{toc}{section}{References}
\bibliographystyle{apalike}

{\scriptsize \bibliography{../../Biblio}}
\end{multicols}

\end{document}